\documentclass[pra,twocolumn,superscriptaddress,10pt]{revtex4-1}
\usepackage{graphicx}
\usepackage{dcolumn}
\usepackage{color}
\usepackage{times}
\usepackage{bm}
\usepackage{amssymb}
\usepackage{amsmath}
\usepackage{epsfig}
\usepackage{epstopdf}
\usepackage{dsfont}
\usepackage{subfigure}
\usepackage{multirow}
\usepackage[colorlinks=true,citecolor=blue, linkcolor=blue]{hyperref}%
\begin{document}
	
	\title{\textbf{Experimental demonstration of complementarity relations between quantum steering criteria}}
	
	\author{Huan Yang}
	\affiliation{School of Physics and Material Science, Anhui University, Hefei 230601, China}
	\affiliation{Institutes of Physical Science and Information Technology, Anhui University, Hefei 230601, China}
	\affiliation{Department of Experiment and Practical Training Management, West Anhui University, Lu’an 237012, China}

	\author{Zhi-Yong Ding}
	\affiliation{School of Physics and Material Science, Anhui University, Hefei 230601, China}
	\affiliation{School of Physics and Electronic Engineering, Fuyang Normal University, Fuyang 236037, China}
	\affiliation{Key Laboratory of Functional Materials and Devices for Informatics of Anhui Educational Institutions, Fuyang Normal University, Fuyang 236037, China}
	
	\author{Xue-Ke Song}
	\affiliation{School of Physics and Material Science, Anhui University, Hefei 230601, China}
	
	\author{Hao Yuan}
	\affiliation{School of Physics and Material Science, Anhui University, Hefei 230601, China}
	\affiliation{CAS Key Laboratory of Quantum Information, University of Science and Technology of China, Hefei 230026, China}	
	\affiliation{Key Laboratory of Opto-Electronic Information Acquisition and Manipulation of Ministry of Education, Anhui University, Hefei 230601, China}
	
	\author{Dong Wang}
	\affiliation{School of Physics and Material Science, Anhui University, Hefei 230601, China}
	\affiliation{CAS Key Laboratory of Quantum Information, University of Science and Technology of China, Hefei 230026, China}	
	
	\author{Jie Yang}
	\affiliation{School of Physics and Material Science, Anhui University, Hefei 230601, China}
	
	\author{Chang-Jin Zhang}
	\affiliation{School of Physics and Material Science, Anhui University, Hefei 230601, China}
	\affiliation{Institutes of Physical Science and Information Technology, Anhui University, Hefei 230601, China}
	
	\author{Liu Ye}
	\email[]{yeliu@ahu.edu.cn}
	\affiliation{School of Physics and Material Science, Anhui University, Hefei 230601, China}
	
	\begin{abstract}
		The ability that one system immediately affects another one by using local measurements is regarded as quantum steering, which can be detected by various steering criteria. Recently, Mondal \textit{et al}. [Phys. Rev. A 98, 052330 (2018)] derived the complementarity relations of coherence steering criteria, and revealed that the quantum steering of system can be observed through the average coherence of subsystem. Here, we experimentally verify the complementarity relations between quantum steering criteria by employing two-photon Bell-like states and three Pauli operators. The results demonstrate that if prepared quantum states can violate two setting coherence steering criteria and turn out to be steerable states, then it cannot violate the complementary settings criteria. Three measurement settings inequality, which establish a complementarity relation between these two coherence steering criteria, always holds in experiment. Besides, we experimentally certify that the strengths of coherence steering criteria dependent on the choice of coherence measure. In comparison with two setting coherence steering criteria based on ${l_1}$ norm of coherence and relative entropy of coherence, our experimental results show that the steering criterion based on skew information of coherence is more stronger in detecting the steerability of quantum states. Thus, our experimental demonstrations can deepen the understanding of the relation between the quantum steering and quantum coherence. 
	\end{abstract}
	
	\maketitle
	
	\section{INTRODUCTION}
		Quantum steering describes a nontrivial trait of quantum world that one subsystem of bipartite systems can instantaneously affects another one by using local measurements \cite{w01, w02}. In contrast to the entanglement \cite{w03} and Bell nonlocality \cite{w04, w05}, quantum steering has been attracted extensive attentions in the field of quantum information only in recent years \cite{w06}. The detection of quantum steering can be realized through the violations of steering criteria (also called steering inequalities), which can be obtained by using correlations, state assemblages, and full information \cite{w07}. There are various steering criteria, including linear and nonlinear steering criteria \cite{w08, w09}, steering inequality from uncertainty relations \cite{w10, w11, w12, w13}, steering criterion from geometric Bell-like inequality \cite{w14}, steering inequalities from the semidefinite programs \cite{w15}, and full information steering inequality \cite{w16}. So far, quantum steering embodies vital application values in subchannel discrimination, resource theory of steering, quantum communication, quantum teleportation, randomness generation, and so on \cite{w07}. Also, it has been demonstrated in a series of significant experiments \cite{w17, w18, w19, w20, w21, w22, w23, w24, w25, w26}.
				
		Coherence, which originates from the superposition principle of quantum mechanics, reflects one of fundamental essences in many quantum phenomena \cite{w27, w28}. Although the inverstigations concerning coherence have a long-standing history, however, the rigorous quantification of coherence in the field of quantum information had never estalished. Untill 2014, based on incoherent operations, Baumgratz \textit{et al}. \cite{w29} put forward the general frame of quantifing coherence for quantum states, and this quantification relies on a fixed reference basis. One can measure quantum coherence via ${l_1}$ norm of coherence \cite{w29}, relative entropy of coherence \cite{w29}, robustness of coherence \cite{w30, w31}, and skew information of coherence \cite{w32, w33, w34}. Recently, quantum coherence becomes a hot topic in both theory \cite{w35, w36, w37, w38, w39, w40} and experiment \cite{w41, w42, w43, w44}. It plays a central part in different fields, such as quantum metrology, quantum thermodynamics, quantum algorithms, and quantum channel discrimination \cite{w27}.
		
		Noteworthily, it is a new tendency for theoretically exploring the complementarity and trade-off relations among different quantities in recent years. According to these relations, a bound for a quantity can be established via another complementary quantity. The understanding of the quantum state space and information as well as correlation can also deepened through these relations. There are several promising efforts in concerning fields \cite{w45, w46, w47, w48, w49, w50, w51, w52}. Singh \textit{et al}. \cite{w45} obtained complementarity between maximal coherence and mixedness, and examined the limits imposed by mixedness of a quantum system with respect to quantum coherence. Cheng \textit{et al}. \cite{w46} explored complementarity relations between the coherences of mutually unbiased bases, and derived relations among coherence, purity, and uncertainty. Considering the maximal violations of the Clauser-Horne-Shimony-Holt inequality, the trade-off relations of Bell violations among pairwise qubit systems were investigated by Qin \textit{et al}.\cite{w47}, and the relations constrain the distribution of nonlocality among the subsystems . By using the relative entropy of coherence, Sharma \textit{et al}. \cite{w48} presented the trade-off relation between the system’s coherence and disturbance induced by a completely positive trace-preserving map. For a multipartite system, the trade-off relations for tripartite nonlocality were established by Zhao \textit{et al}. \cite{w49}. Experimentally, different complementarity and trade-off relations were also tested \cite{w53, w54, w55}. By employing a photonic qutrit-qubit hybrid system, Zhan \textit{et al}. \cite{w53} experimentally verified contextuality-nonlocality trade-off relation, and the results certified that entanglement is a particular form for fundamental quantum resource. Weston \textit{et al}. \cite{w54} experimentally tested the universally valid complementarity relations satisfied for any joint measurement of two observables . In two noncommuting reference bases, Lv \textit{et al}. \cite{w55} experimentally verified the trade-off relation of quantum coherence, and their results displayed that the lower and upper bounds restrict the sum of quantum coherence under these bases .
		
		Recently, Mondal \textit{et al}. \cite{w56} obtained the complementarity relations between coherence steering criteria by employing different quantifications of quantum coherence. This work established a connection between two valuable quantum resources in quantum information task, i.e., quantum steering and quantum coherence. However, the test of the complementarity relations in experiment is still lacking. The concerning investigation may further deepen our understanding of the relation between the quantum steering and quantum coherence in practice. Also, it can demonstrate a new method to detect quantum steering in experiment, namely, witness quantum steering of system via detecting quantum coherence of subsystem. Motivated by this, we experimentally verify the complementarity relations between different coherence steering criteria in this work. The experimental results show that one setting coherence steering criteria cannot be violated if its complementary criteria can be violated by prepared states. In comparison with two setting coherence steering criteria from ${l_1}$ norm of coherence and relative entropy of coherence, the prepared states are more easy to violate two setting coherence steering criterion based on skew information of coherence, suggesting that it can detect more steerable states in experiment.
		
	\section{COHERENCE STEERING CRITERIA AND COMPLEMENTARITY RELATIONS}
		Considering two quantum systems A and B prepared by Alice, which form an entangled state ${\rho _{AB}}$. And then, the system B is sent to Bob. The task of Alice is to make Bob believe the fact that the state shared by them is indeed entangled, hence, the nonlocal correlation is shared by them. However, Bob does not trust Alice and only considers the system B is quantum. Also, Bob thinks that Alice may use a single quantum system B to cheat him \cite{w17}. If and only if the state of Bob cannot be represented by using the local hidden state (LHS) model \cite{w06} $\rho _a^A = \sum\nolimits_\lambda  {p(\lambda )p(a|A,\lambda )\rho _B^Q(\lambda )} $, then Bob accepts the fact that the ${\rho _{AB}}$ prepared by Alice is an entangled state, and nonlocal correlation is shared by them. Here, $\lambda $  is hidden variable with $\sum\nolimits_\lambda  {p(\lambda )}  = 1$, and $\rho _B^Q(\lambda )$ is LHS. $\{ p(\lambda ),\rho _B^Q(\lambda )\} $ denotes an ensemble of preexisting LHS for Bob, $p(a|A,\lambda )$ indicates the stochastic map of Alice, which is used to convince or fool Bob via $\rho _a^A$. The ${\rho _{AB}}$ is steerable state if and only if the joint probabilities of measurement outcomes (Alice performs the measurement of \textit{A} on her subsystem and obtains the outcome $a \in \{ 0,1\} $, Bob performs the measurement of \textit{B} and obtains the outcome $b \in \{ 0,1\} $) cannot be described by employing a local hidden variable-local hidden state (LHV-LHS) model, namely $P(a,b|A,B,{\rho _{AB}}) = \sum\nolimits_\lambda  {P(\lambda )P(a|A,\lambda ){P_Q}(b|B,{\rho _\lambda })} $, where ${P_Q}(b|B,{\rho _\lambda })$ is probability of the outcome \textit{b} corresponding to measurement \textit{B}.
		
		In 2018, Mondal \textit{et al}. \cite{w56} proposed the coherence steering criteria, which can help us to observe the quantum steering of a two-qubit state via the quantum coherence of subsystem. Consider two-qubit states ${\rho _{AB}} = (\mathbb{I} \otimes \mathbb{I} + {\bf{r}} \cdot {\bf{\sigma }} \otimes \mathbb{I} + \mathbb{I} \otimes {\bf{s}} \cdot {\bf{\sigma }} + \sum\nolimits_{i,j} {{t_{ij}}{\sigma _i} \otimes {\sigma _j}} )/4$ and three Pauli operators ${\rm{\{ }}{\sigma _i}{\rm{\} }}(i = x,y,z)$ as a complete set of mutually unbiased bases (MUBs), where ${\bf{r}} = {\rm{tr}}\left( {{\rho _{AB}}{\bf{\sigma }} \otimes \mathbb{I}} \right)$, ${\bf{s}} = {\rm{tr}}\left( {{\rho _{AB}}\mathbb{I} \otimes {\bf{\sigma }}} \right)$, and ${t_{ij}} = {\text{tr}}\left( {{\rho _{AB}}{\sigma _i} \otimes {\sigma _j}} \right){\text{ }}$. Assuming that Alice implements a projective measurement on her system by using the eigenbasis of ${\sigma _i}$ and the corresponding outcome is $a \in \{ 0,1\} $. The corresponding probability is $p({\rho _{\left. B \right|M_a^i}}) = {\text{tr}}[(M_a^i \otimes \mathbb{I}){\rho _{AB}}]$, and measurement operator is $M_a^i = [\mathbb{I} + {( - 1)^a}{\sigma _i}]/2$. Similarly, Alice can measure her system by employing another measurement operator $M_a^k = [\mathbb{I} + {( - 1)^a}{\sigma _k}]/2$, and $k \ne i,k \in \{ x,y,z\} $. For each projective measurement implemented by Alice, Bob can measure the coherence of his conditional state ${\rho _{\left. B \right|M_a^i}}$ under the eigenbasis of every Pauli operator. According to the number of Pauli operators chosen by Bob, the coherence steering criteria can be divided into one measurement setting (or one setting) and two measurement setting (or two setting), respectively \cite{w56}. Explicitly, the probability superposition of coherences of ${\rho _{\left. B \right|M_a^i}}$  can be defined as		
		
		\begin{equation}\label{g01}
			S_\ell ^B({\rho _{AB}}) = \sum\limits_{i,a} {p({\rho _{\left. B \right|M_a^i}})C_{i + \ell }^q({\rho _{\left. B \right|M_a^i}})},
		\end{equation}		
		where $\ell  \in \{ 0,1,2\} $, $C_i^q(\rho )$ represents different coherence measures under the eigenbasis of Pauli operator ${\sigma _i}$, including ${l_1}$ norm of coherence ($q = {l_1}{\text{C}}$ and $C_i^{{l_1}{\text{C}}}(\rho ) = \sum\nolimits_{i \ne j} {\left| {\left\langle {{k_i}} \right|\rho \left| {{k_j}} \right\rangle } \right|} $), relative entropy of coherence ($q=\text{REC}$ and $C_i^\text{REC}(\rho ) = S({\rho ^{diag}}) - S(\rho )$, and skew information of coherence ($q = \text{SIC}$ and $C_i^\text{SIC}(\rho ) =  - ({\text{Tr[}}\sqrt \rho  ,{\sigma _i}{{\text{]}}^2})/2$). $\{ \left| {{k_i}} \right\rangle ,\left| {{k_j}} \right\rangle \}$ denote the eigenvectors of ${\sigma _i}$, $S(x)$ is the von Neumann entropy, and ${\rho ^{diag}}{\text{ = }}\sum\nolimits_i {\left| {{k_i}} \right\rangle \left\langle {{k_i}} \right|\rho \left| {{k_i}} \right\rangle } \left\langle {{k_i}} \right|$. If Bob measures the coherence only in one Pauli operator ${\sigma _i}$, which is same as each projective measurement chosen by Alice (i.e., $\ell  = 0$), and then the one setting coherence steering criteria are given by
				
		\begin{equation}\label{g02}
			S_0^B({\rho _{AB}}) = \sum\limits_{i,a} {p({\rho _{\left. B \right|M_a^i}})C_i^q({\rho _{\left. B \right|M_a^i}})}  \leqslant {\varepsilon ^q}.
		\end{equation}
		Here, $q \in \{ {l_1}\text{C},\text{REC},\text{SIC}\}$, and ${\varepsilon ^q} \in \{ \sqrt 6 ,2.23,2\} $ represent corresponding upper bounds for different coherence measures. The criteria cannot be violated by the state with the LHS model. If Bob measures the coherence in the eigenbasis of another two Pauli operators ${\sigma _j}$ and ${\sigma _k}$ ($j \ne k \ne i$) (corresponding to $\ell  = 1,2$), which are different from the one chosen by Alice's measurement, there exists the two setting coherence steering criteria 

		\begin{equation}\label{g03}
			\frac{1}{2}S_{12}^B({\rho _{AB}}) = \frac{1}{2}(S_1^B({\rho _{AB}}) + S_2^B({\rho _{AB}})) \leqslant {\varepsilon ^q}.
		\end{equation}
		 Any state with the LHS model obeys these steering criteria. The violation of the criteria means that the ${\rho _{AB}}$ is steerable. If Bob measures the coherence under the eigenbasis of three Pauli operators after each projective measurement performed by Alice on her subsystem, the inequality of three measurement settings is
		
		\begin{equation}\label{g04}
			\frac{1}{3}S_{012}^B({\rho _{AB}}) = \frac{1}{3}(S_0^B({\rho _{AB}}) + S_1^B({\rho _{AB}}) + S_2^B({\rho _{AB}})) \leqslant {\varepsilon ^q}.
		\end{equation}		
		This inequality cannot be used to detect the quantum steering of ${\rho _{AB}}$ due to the fact that the inequality is satisfied by all two-qubit states. In reality, the Eq. (4) describes a complementarity relation between one setting coherence steering criteria and two setting coherence steering criteria, that is,
		\begin{equation}\label{g05}
			\frac{1}{3}S_{012}^B({\rho _{AB}}) = \frac{1}{3}(S_0^B({\rho _{AB}}) + S_{12}^B({\rho _{AB}})) \leqslant {\varepsilon ^q}.
		\end{equation}		
		The results manifest that if one criterion between Eqs. (2) and (3) is violated, and then the other one as a compensation can never be violated.
		
	\section{EXPERIMENTAL DEMONSTRATIONS AND RESULTS}
		In the process of our experimental implementation, we choose two-photon Bell-like states as test states. The polarized photons are encoded as qubits. The horizontally and vertically polarized states are described by using $\left| H \right\rangle $ and $\left| V \right\rangle $, respectively. Hence, two-photon Bell-like states are

		\begin{equation}\label{g05}
		{\rho _{{\text{AB}}}} = \left| {{\phi _{AB}}} \right\rangle \left\langle {{\phi _{AB}}} \right|
		\end{equation}			 
		 with		 
		\begin{equation}\label{g06}
		\left| {{\phi _{AB}}} \right\rangle {\text{ = }}\cos \theta \left| {HH} \right\rangle  + \sin \theta \left| {VV} \right\rangle.
		\end{equation}
		
		\begin{figure}
			\centering
			\includegraphics[width=8cm]{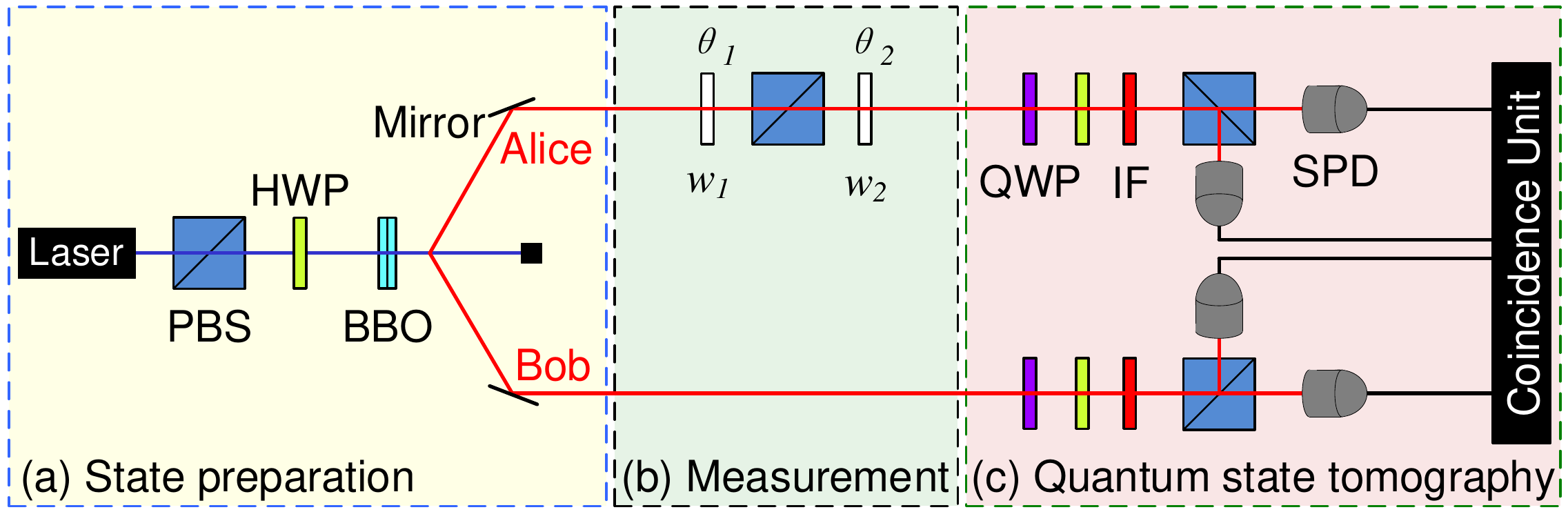}\\
			\caption{Experimental setup. The module (a) is used to prepared two-photon Bell-like states. Module (b) is used to achieve local measurement on photon of Alice and the prepared states will be transformed into the postmeasurement state. Module (c) is used to carry out three Pauli measurement for the photons of Alice and Bob, and also perform the tomography for quantum state. Abbreviations: PBS, polarizing beam splitter; HWP, half-wave plate; BBO, type-I $\beta $-barium borate; QWP, quarter-wave plate; IF: interference filter; SPD: single photon detector. ${w_1}$ and ${w_2}$: the types of wave plates; ${\theta _1}$ and ${\theta _2}$: the angles of optical axes of wave plates.}\label{Fig1}
		\end{figure}		
		
		\begin{table}[b]
			\centering
			\caption{The settings of wave plates for realizing different PMOs on the photon of Alice in the module (b).}
			\begin{ruledtabular}
				\begin{tabular}{ccccccc}
					Settings      & ${\text{M}}_0^x$  & ${\text{M}}_1^x$ & ${\text{M}}_0^y$ & ${\text{M}}_1^y$ & ${\text{M}}_0^z$ & ${\text{M}}_1^z$ \\
					\colrule
					${w_1}$       &        HWP        &        HWP       &        QWP       &        QWP       &        HWP       &        HWP       \\
					${\theta _1}$ & $ 22.5 ^{\circ}$  & $-22.5 ^{\circ}$ & $45 ^{\circ}$    & $-45 ^{\circ}$   & $ 0 ^{\circ}$    & $45 ^{\circ}$    \\
					${w_2}$       &        HWP        &        HWP       &        QWP       &        QWP       &        HWP       &        HWP       \\
					${\theta _2}$ & $ 22.5 ^{\circ}$  & $-22.5 ^{\circ}$ & $-45 ^{\circ}$   & $45 ^{\circ}$    & $ 0 ^{\circ}$    & $45 ^{\circ}$    \\
				\end{tabular}
			\end{ruledtabular}		
		\end{table}
	
		Figure 1 provides the schematic diagram of all-optical experiment setup which is used to realize the verification of the complementarity relations. The setup contains three modules. The yellow area is the module (a) to prepare test states. To be explicit, high-power continuous pumped beam (the power is 130mW and the wavelength is 405nm) passes through the polarization beam splitter (PBS). The state of pumped beam transforms into horizontally polarized state $\left| H \right\rangle $. This light beam first passes through the half-wave plate (HWP), and then is focused on two type-I $\beta $-barium borate (BBO) crystals ($6.0 \times 6.0 \times 0.5{\text{m}}{{\text{m}}^3}$, and the optical axis is cut at ${29.2^{\text{o}}}$). Under the spontaneous parametric down conversion \cite{w57}, Bell-like states $\left| {{\phi _{AB}}} \right\rangle {\text{ = }}\cos \theta \left| {HH} \right\rangle  + \sin \theta \left| {VV} \right\rangle $ shared by a pair of entangled photons (the central wavelength is 810 nm) are prepared. The state parameter $\theta $ can be easily changed by controlling the angle of optical axis of HWP. Experimentally, we set $\theta $ to ${0^{\text{o}}}$, ${10^{\text{o}}}$, ${20^{\text{o}}}$, ${30^{\text{o}}}$, ${40^{\text{o}}}$, ${45^{\text{o}}}$, ${50^{\text{o}}}$, ${60^{\text{o}}}$, ${70^{\text{o}}}$, ${80^{\text{o}}}$, and ${90^{\text{o}}}$, respectively. The average fidelity of these test states is $\bar F = 0.9987 \pm 0.0041$, which is obtained according to $F(\rho ,{\rho _0}) \equiv {\text{Tr}}\sqrt {\sqrt \rho  {\rho _0}\sqrt \rho  } $. Here, the theoretical and experimental density matrices are indicated by ${\rho _0}$ and $\rho $, respectively. In experiment, we estimate the error bars according to the statistical variation of coincidence counts obeyed the Poisson distribution \cite{w58}. The green area in Fig. 1 is the module (b) of local projective measurement, and the module is to implement the local projective measurement on the photon of Alice. This module consists of a PBS and two wave plates (denoted by ${w_1}$ and ${w_2}$, corresponding rotation angles of the optical axes are ${\theta _1}$ and ${\theta _2}$). A complete set of MUB measurements are $X = \{ \left| D \right\rangle \left\langle D \right|,\left| A \right\rangle \left\langle A \right|\} $, $Y = \{ \left| R \right\rangle \left\langle R \right|,\left| L \right\rangle \left\langle L \right|\} $, and $Z = \{ \left| H \right\rangle \left\langle H \right|,\left| V \right\rangle \left\langle V \right|\} $. Here, $\{ \left| D \right\rangle {\text{ = }}(\left| H \right\rangle  + \left| V \right\rangle )/\sqrt 2 ,\left| A \right\rangle {\text{ = }}(\left| H \right\rangle  - \left| V \right\rangle )/\sqrt 2 \} $, $\{ \left| R \right\rangle {\text{ = }}(\left| H \right\rangle  + i\left| V \right\rangle )/\sqrt 2 ,\left| L \right\rangle {\text{ = }}(\left| H \right\rangle  - i\left| V \right\rangle )/\sqrt 2 \} $, and $\{ \left| H \right\rangle ,\left| V \right\rangle \} $ are eigenvectors of ${\sigma _x}$, ${\sigma _y}$, and ${\sigma _z}$, respectively. For simplicity, we use ${\text{M}}_0^x{\text{ = }}\left| D \right\rangle \left\langle D \right|$, ${\text{M}}_1^x{\text{ = }}\left| A \right\rangle \left\langle A \right|$, ${\text{M}}_0^y = \left| R \right\rangle \left\langle R \right|$, ${\text{M}}_1^y = \left| L \right\rangle \left\langle L \right|$, ${\text{M}}_0^z = \left| H \right\rangle \left\langle H \right|$, and ${\text{M}}_1^z = \left| V \right\rangle \left\langle V \right|$ to represent six projection measurement operators (PMOs) in our experiment. These operators can be realized by adjusting the types and angles of wave plates in module (b), as shown in Table I. For each Bell-like state prepared in experiment, the postmeasurement states are expressed by ${\rho _{{\text{M}}_0^xB}}$, ${\rho _{{\text{M}}_1^xB}}$, ${\rho _{{\text{M}}_0^yB}}$, ${\rho _{{\text{M}}_1^yB}}$, ${\rho _{{\text{M}}_0^zB}}$, and ${\rho _{{\text{M}}_1^zB}}$. The pink area in Fig. 1 indicates the module (c) of the quantum state tomography \cite{w59, w60}, which is used to attain density matrices of quantum states.

		\begin{figure}
			\centering
			\includegraphics[width=8cm]{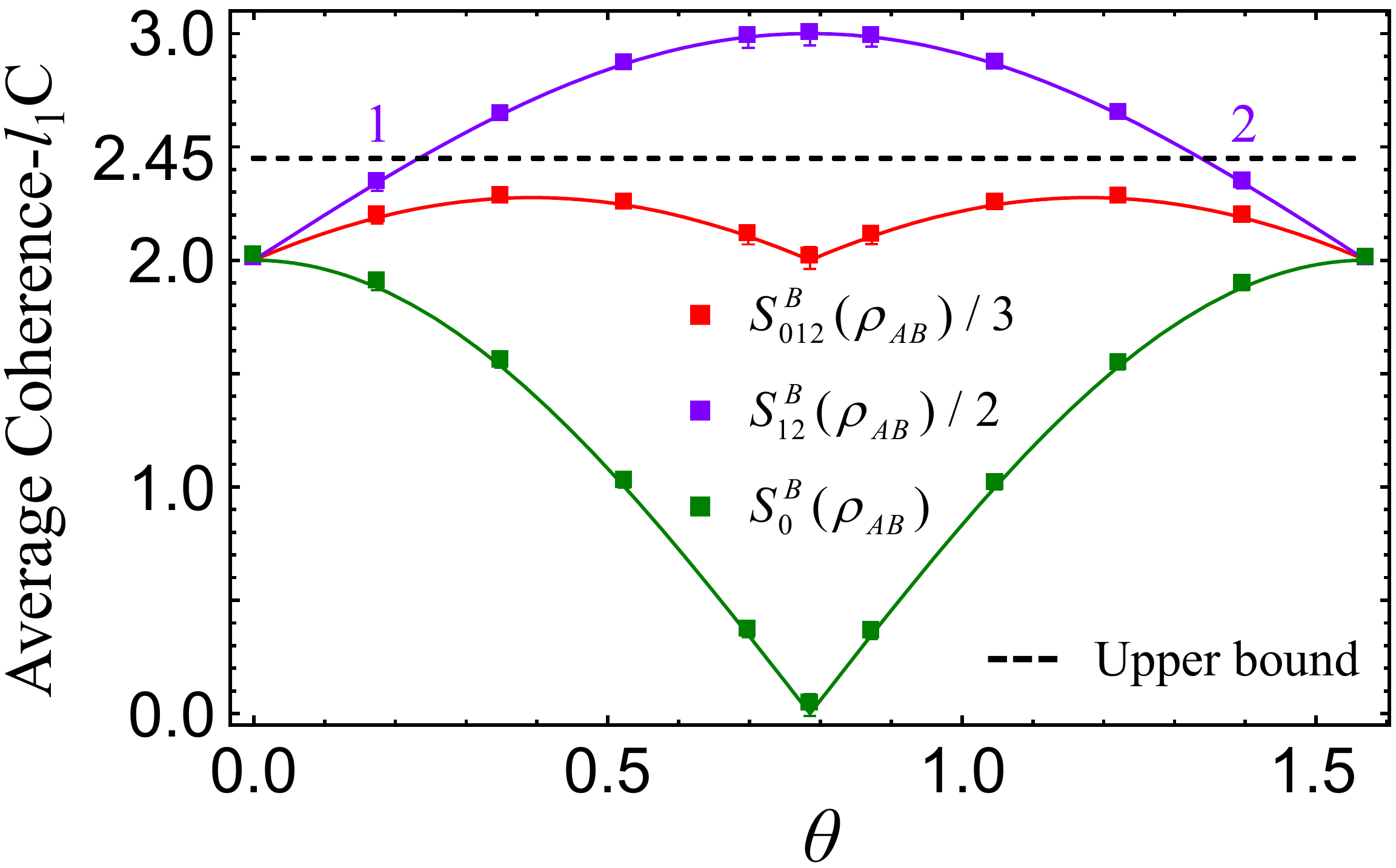}\\
			\caption{Experimental results of the ${l_1}$ norm of coherence. The red squares indicate the experimental results of $S_{012}^B({\rho _{AB}})/3$. The purple squares denote the experimental results of $S_{12}^B({\rho _{AB}})/2$ and the green squares represent the experimental results of $S_0^B({\rho _{AB}})$, respectively. The corresponding theoretical predictions are represented by using solid lines with different colors. The black dashed line represents the upper bound ${\varepsilon ^{{l_1}\text{C}}} = \sqrt 6 $.}\label{Fig2}
		\end{figure}
		
		\begin{figure}
			\centering
			\includegraphics[width=8cm]{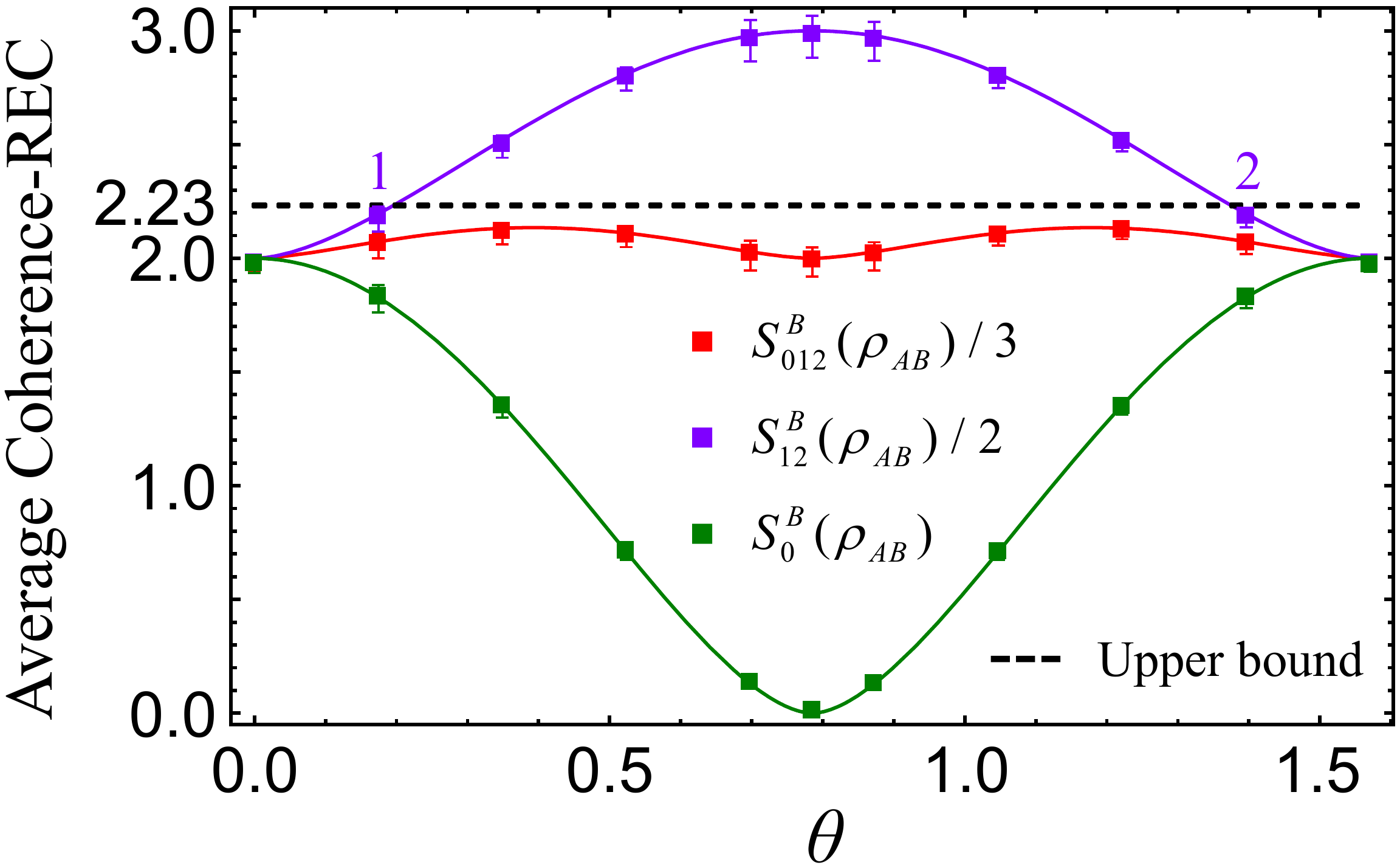}\\
			\caption{Experimental results of the relative entropy of coherence. The red squares indicate the experimental results of $S_{012}^B({\rho _{AB}})/3$. The purple squares denote the experimental results of $S_{12}^B({\rho _{AB}})/2$ and the green squares represent the experimental results of $S_0^B({\rho _{AB}})$, respectively. The corresponding theoretical predictions are represented by employing solid lines with different colors. The black dashed line represents the upper bound ${\varepsilon ^\text{REC}} = 2.23$.}\label{Fig3}
		\end{figure}
	
		\begin{figure}
			\centering
			\includegraphics[width=8cm]{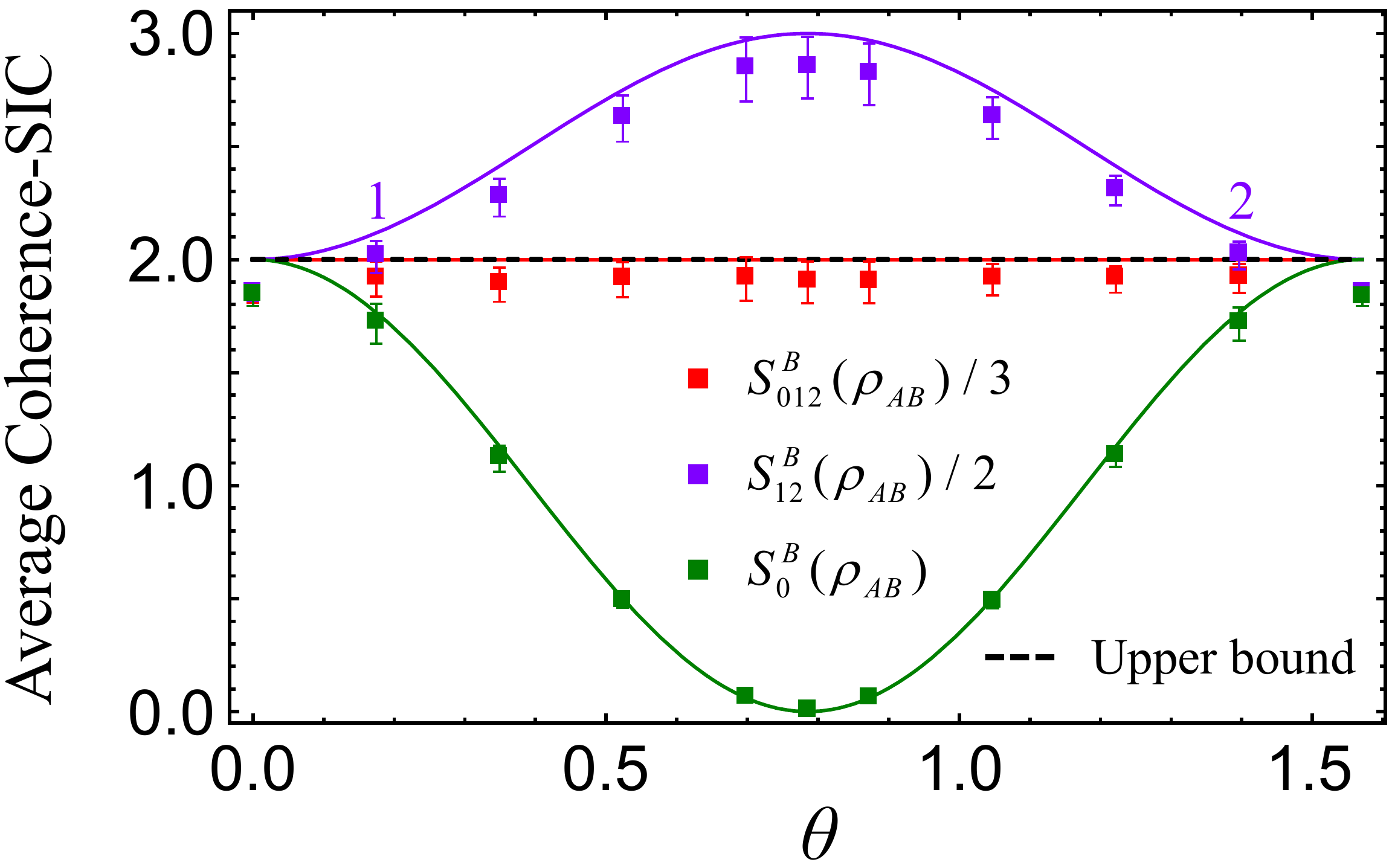}\\
			\caption{Experimental results of skew information of coherence. The red squares indicate the experimental results of $S_{012}^B({\rho _{AB}})/3$. The purple squares denote the experimental results of $S_{12}^B({\rho _{AB}})/2$ and the green squares represent the experimental results of $S_0^B({\rho _{AB}})$, respectively. the corresponding theoretical predictions are represented by using solid lines with different colors. The black dashed line represents the upper bound ${\varepsilon ^\text{SIC}} = 2$.}\label{Fig4}
		\end{figure}

		\begin{table}
			\centering
			\caption{Experimental data of $S_{012}^B({\rho _{AB}})/3$, $S_{12}^B({\rho _{AB}})/2$, and $S_0^B({\rho _{AB}})$ for the ${l_1}$ norm of coherence.}
			\begin{ruledtabular}
				\begin{tabular}{ccccc}
					$\theta$ & $S_{012}^B({\rho _{AB}})/3$ & $S_{12}^B({\rho _{AB}})/2$ & $S_0^B({\rho _{AB}})$\\
					\colrule
					$0^{\text{o}}$ & 2.0062$\pm $0.0195 & 2.0017$\pm $0.0200 & 2.0150$\pm $0.0186 \\
					$10^{\text{o}}$& 2.1915$\pm $0.0307 & 2.3376$\pm $0.0303 & 1.8993$\pm $0.0317 \\					
					$20^{\text{o}}$& 2.2750$\pm $0.0197 & 2.6373$\pm $0.0171 & 1.5504$\pm $0.0250 \\					
					$30^{\text{o}}$& 2.2474$\pm $0.0171 & 2.8617$\pm $0.0149 & 1.0188$\pm $0.0215 \\					
					$40^{\text{o}}$& 2.1085$\pm $0.0389 & 2.9810$\pm $0.0447 & 0.3636$\pm $0.0274 \\					
					$45^{\text{o}}$& 2.0095$\pm $0.0474 & 2.9954$\pm $0.0465 & 0.0378$\pm $0.0493 \\					
					$50^{\text{o}}$& 2.1060$\pm $0.0356 & 2.9808$\pm $0.0389 & 0.3562$\pm $0.0289 \\					
					$60^{\text{o}}$& 2.2453$\pm $0.0150 & 2.8628$\pm $0.0117 & 1.0102$\pm $0.0216 \\					
					$70^{\text{o}}$& 2.2740$\pm $0.0138 & 2.6423$\pm $0.0113 & 1.5372$\pm $0.0189 \\					
					$80^{\text{o}}$& 2.1892$\pm $0.0207 & 2.3396$\pm $0.0214 & 1.8884$\pm $0.0194 \\					
					$90^{\text{o}}$& 2.0018$\pm $0.0153 & 2.0008$\pm $0.0156 & 2.0038$\pm $0.0148 \\
				\end{tabular}
			\end{ruledtabular}		
		\end{table}	
	
		\begin{table}
			\centering
			\caption{Experimental data of $S_{012}^B({\rho _{AB}})/3$, $S_{12}^B({\rho _{AB}})/2$, and $S_0^B({\rho _{AB}})$ for the relative entropy of coherence.}
			\begin{ruledtabular}
				\begin{tabular}{ccccc}
					$\theta$ & $S_{012}^B({\rho _{AB}})/3$ & $S_{12}^B({\rho _{AB}})/2$ & $S_0^B({\rho _{AB}})$\\
					\colrule
					$0^{\text{o}}$ & 1.9680$\pm $0.0238 & 1.9686$\pm $0.0199 & 1.9668$\pm $0.0315 \\					
					$10^{\text{o}}$& 2.0557$\pm $0.0565 & 2.1729$\pm $0.0550 & 1.8213$\pm $0.0596 \\										
					$20^{\text{o}}$& 2.1079$\pm $0.0463 & 2.4911$\pm $0.0490 & 1.3416$\pm $0.0409 \\										
					$30^{\text{o}}$& 2.0934$\pm $0.0442 & 2.7880$\pm $0.0512 & 0.7043$\pm $0.0303 \\										
					$40^{\text{o}}$& 2.0127$\pm $0.0654 & 2.9562$\pm $0.0911 & 0.1257$\pm $0.0141 \\										
					$45^{\text{o}}$& 1.9832$\pm $0.0649 & 2.9741$\pm $0.0920 & 0.0015$\pm $0.0107 \\										
					$50^{\text{o}}$& 2.0089$\pm $0.0618 & 2.9531$\pm $0.0859 & 0.1204$\pm $0.0136 \\										
					$60^{\text{o}}$& 2.0932$\pm $0.0369 & 2.7898$\pm $0.0421 & 0.7000$\pm $0.0264 \\										
					$70^{\text{o}}$& 2.1149$\pm $0.0309 & 2.5030$\pm $0.0334 & 1.3388$\pm $0.0258 \\										
					$80^{\text{o}}$& 2.0570$\pm $0.0393 & 2.1761$\pm $0.0402 & 1.8187$\pm $0.0375 \\										
					$90^{\text{o}}$& 1.9658$\pm $0.0152 & 1.9681$\pm $0.0126 & 1.9613$\pm $0.0204 \\
				\end{tabular}
			\end{ruledtabular}		
		\end{table}	
	
		\begin{table}
			\centering
			\caption{Experimental data of $S_{012}^B({\rho _{AB}})/3$, $S_{12}^B({\rho _{AB}})/2$, and $S_0^B({\rho _{AB}})$ for skew information of coherence.}
			\begin{ruledtabular}
				\begin{tabular}{ccccc}
					$\theta$ & $S_{012}^B({\rho _{AB}})/3$ & $S_{12}^B({\rho _{AB}})/2$ & $S_0^B({\rho _{AB}})$\\
					\colrule
					$0^{\text{o}}$ & 1.8462$\pm $0.0364 & 1.8485$\pm $0.0312 & 1.8415$\pm $0.0469 \\										
					$10^{\text{o}}$& 1.9129$\pm $0.0773 & 2.0112$\pm $0.0717 & 1.7164$\pm $0.0887 \\															
					$20^{\text{o}}$& 1.8884$\pm $0.0751 & 2.2734$\pm $0.0835 & 1.1184$\pm $0.0581 \\															
					$30^{\text{o}}$& 1.9113$\pm $0.0775 & 2.6241$\pm $0.1025 & 0.4857$\pm $0.0275 \\															
					$40^{\text{o}}$& 1.9139$\pm $0.0962 & 2.8415$\pm $0.1412 & 0.0587$\pm $0.0061 \\															
					$45^{\text{o}}$& 1.8994$\pm $0.0924 & 2.8488$\pm $0.1353 & 0.0005$\pm $0.0066 \\															
					$50^{\text{o}}$& 1.8982$\pm $0.0922 & 2.8193$\pm $0.1352 & 0.0560$\pm $0.0064 \\															
					$60^{\text{o}}$& 1.9117$\pm $0.0695 & 2.6266$\pm $0.0920 & 0.4819$\pm $0.0247 \\															
					$70^{\text{o}}$& 1.9118$\pm $0.0584 & 2.3041$\pm $0.0656 & 1.1273$\pm $0.0441 \\															
					$80^{\text{o}}$& 1.9166$\pm $0.0647 & 2.0180$\pm $0.0607 & 1.7139$\pm $0.0727 \\															
					$90^{\text{o}}$& 1.8413$\pm $0.0264 & 1.8475$\pm $0.0222 & 1.8288$\pm $0.0346 \\
				\end{tabular}
			\end{ruledtabular}		
		\end{table}	
	
		 \begin{table}[h]
			\centering
			\caption{The settings of HWP and QWP in the module (c) for different PMOs implemented on both Alice’s photon and Bob’s photon.}
			\begin{ruledtabular}
				\begin{tabular}{ccccccc}
					Settings         & ${\text{M}}_0^x$& ${\text{M}}_1^x$& ${\text{M}}_0^y$& ${\text{M}}_1^y$& ${\text{M}}_0^z$& ${\text{M}}_1^z$ \\
					\colrule
					The angle of HWP & $ 22.5 ^{\circ}$& $-22.5 ^{\circ}$& $22.5 ^{\circ}$ & $-22.5 ^{\circ}$&  $ 0 ^{\circ}$  & $45 ^{\circ}$    \\
					The angle of QWP & $ 45 ^{\circ}$  &  $45 ^{\circ}$  &  $0 ^{\circ}$   &   $0 ^{\circ}$  &  $ 0 ^{\circ}$  & $0 ^{\circ}$    \\
				\end{tabular}
			\end{ruledtabular}		
		\end{table}
	
		\begin{figure}[h]
			\centering
			\includegraphics[width=8cm]{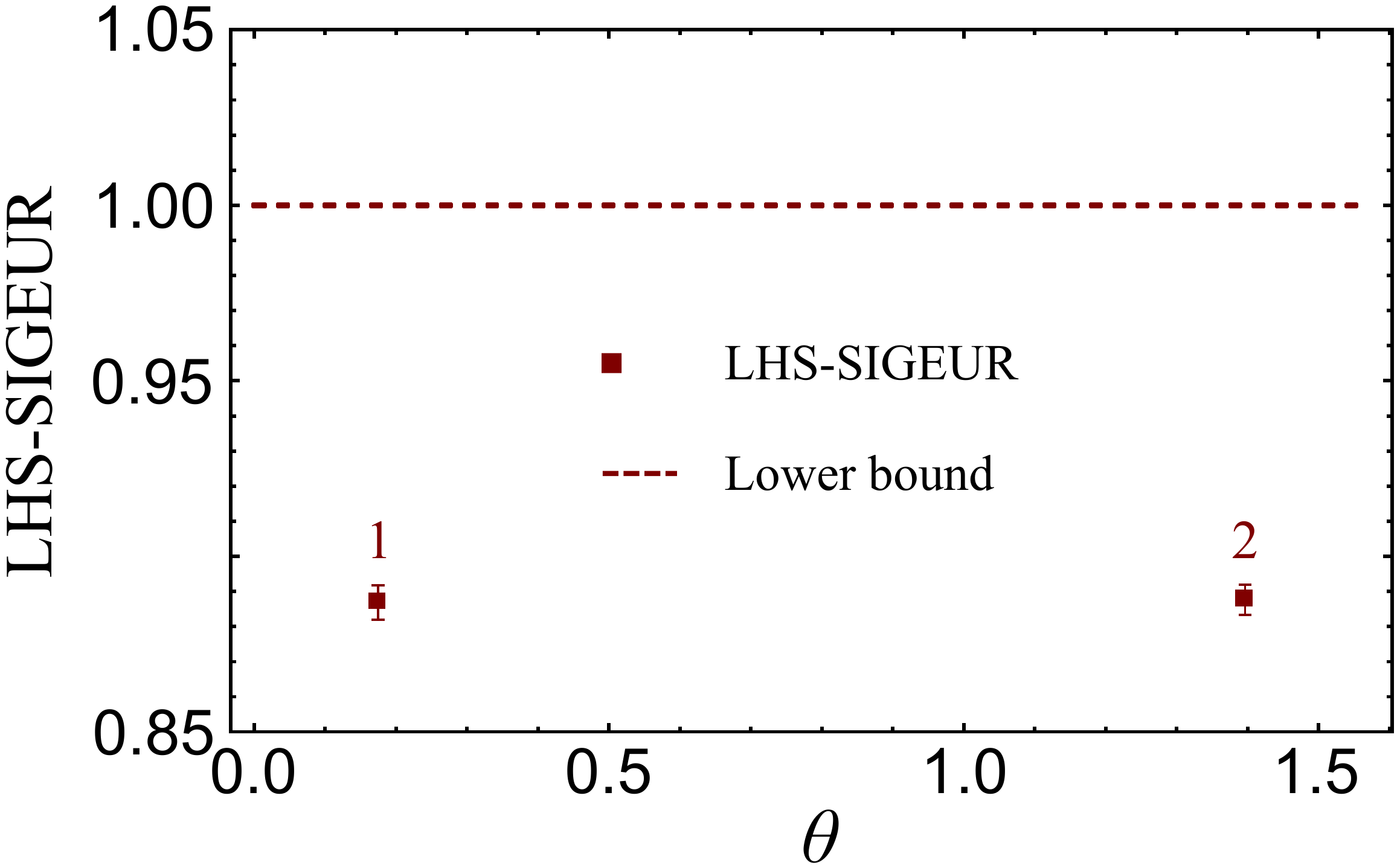}\\
			\caption{Experimental results of LHS-SIGEUR for prepared Bell-like states with $\theta {\text{ = }}{10^{\text{o}}}$ and $\theta {\text{ = }}{80^{\text{o}}}$.}\label{Fig5}
		\end{figure}

		Now let us turn to verify the complementarity relations between quantum steering criteria in experiment. The experimental measurement probabilities $p({\rho _{\left. B \right|M_a^i}})$ in Eqs. (2)-(4) are obtained by virtue of coincidence counts \cite{w61}, and the corresponding coherence $C_i^q({\rho _{\left. B \right|M_a^i}})$ are calculated according to the density matrices of 66 postmeasurement states restructured via quantum state tomography. Thus, the experimental results of $S_0^B({\rho _{AB}})$, $S_{12}^B({\rho _{AB}})/2$, and $S_{012}^B({\rho _{AB}})/3$ can be attained in different coherence measures. In detail, Fig. 2 and Table II depict the results based on ${l_1}$ norm of coherence (${l_1}$C). Figure. 3 and Table III provide the results based on relative entropy of coherence (REC). The results based on skew information of coherence (SIC) are depicted in Fig. 4 and Table IV. It is worthwhile to note that some of the error bars are too short to exhibit in Fig. 2-4. For all figures, the green squares, purple squares, and red squares denote the experimental results of  $S_0^B({\rho _{AB}})$, $S_{12}^B({\rho _{AB}})/2$, and $S_{012}^B({\rho _{AB}})/3$, respectively. The corresponding theoretical results are displayed by means of different colored curves, and the black dotted lines are the upper bounds ${\varepsilon ^q}(q = {l_1}\text{C},\text{REC},\text{SIC})$ in Eqs. (2)-(4).   

		As displayed from Figs. 2-4, the experimental results are in good agreement with the theoretical predictions. No matter what kind of coherence measure is chosed, the experimental results of $S_0^B({\rho _{AB}})$ are anticorrelated with the ones of $S_{12}^B({\rho _{AB}})/2$. Moreover, the experimental results reveal that if two setting coherence steering criteria are violated ($S_{12}^B({\rho _{AB}})/2 > {\varepsilon ^q}$) by prepared Bell-like states, one setting coherence steering criteria as compensations cannot be violated ($S_0^B({\rho _{AB}}) \leqslant {\varepsilon ^q}$). All prepared Bell-like states cannot violate three measurement settings inequality ($S_{012}^B({\rho _{AB}})/3 \leqslant {\varepsilon ^q}$).
		
		From Figs. 2-4 and Tables II-IV, one can find that seven prepared Bell-like states ($\theta {\text{ = }}{20^{\text{o}}},{30^{\text{o}}},{40^{\text{o}}},{45^{\text{o}}},{50^{\text{o}}},{60^{\text{o}}},{70^{\text{o}}}$) can violate two setting coherence steering criteria based on different coherence measures (i.e., ${l_1}$C, REC, and SIC). The results demonstrate that these seven prepared Bell-like states are steerable states. It also deserves emphasizing that two prepared Bell-like states with $\theta {\text{ = }}{10^{\text{o}}}$ and $\theta {\text{ = }}{80^{\text{o}}}$ (labeled by 1 and 2 in Figs. 2-4, respectively) can violate two setting coherence steering criterion based on skew information of coherence. However, these two states cannot violate the criteria based on ${l_1}$ norm of coherence and relative entropy of coherence. The experimental results verify that the quantum steering of these two states can only be detected by the two setting coherence steering criterion from skew information of coherence, and cannot be captured through the ones from ${l_1}$  norm of coherence and relative entropy of coherence. In order to further certify the results, we perform steering inequality tests on these two states by using steering inequality from general entropic uncertainty relation (SIGEUR), which is an effective tool to detect quantum steering \cite{w12, w25, w26}. The SIGEUR is written as ${(n - 1)^{ - 1}}\sum\nolimits_i {\{ 1 - \sum\nolimits_{ab} {[{{(p_{ab}^{(i)})}^n}/{{(p_a^{(i)})}^{n - 1}}]} \} }  \geqslant C_B^{(n)}$  with $C_B^{(n)} = m{\ln _n}[md/(d + m - 1)]$ for $n \in (0,2]$. Here, ${\text{l}}{{\text{n}}_n}(x) = ({x^{1 - n}} - 1)/(1 - n)$, and $p_{ab}^{(i)}$ $(i = x,y,z{\text{ and }}a,b \in \{ 0,1\} )$ represents the probability of outcome $(a,b)$ for a set of measurements ${A_i} \otimes {B_i}$ implemented both on the photons of Alice and Bob. $p_a^{(i)}$ is the probability of marginal outcome for measurement ${A_i}$ of Alice. \textit{d} is the dimension of system, and \textit{m} is number of MUBs. In our experiment, $d{\text{ = }}2$, $m{\text{ = }}3$, and we choose $n{\text{ = }}2$ due to that the SIGEUR is the strongest one in this case \cite{w12, w25, w26}. Hence, the lower bound $C_B^{(n)}{\text{ = }}1$. In technology, we remove the module (b) in Fig. 2, and use module (c) to achieve the six PMOs performed on both Alice’s photon and Bob’s photon, as illuminated in Table V. Thereby, the steering inequality test can be implemented on prepared Bell-like states with $\theta {\text{ = }}{10^{\text{o}}}$ and $\theta {\text{ = }}{80^{\text{o}}}$. One can conveniently calculate $p_{ab}^{(i)}$ and $p_a^{(i)}$ according to the coincidence counts in experiment. The experimental results are plotted in Fig. 5. It is demonstrated that the experimental left hand sides of SIGEUR (LHS-SIGEUR) for prepared Bell-like states with $\theta {\text{ = }}{10^{\text{o}}}$ and $\theta {\text{ = }}{80^{\text{o}}}$ are equal to $0.8869 \pm 0.0049$ and $0.8876 \pm 0.0043$, respectively. These results violate the SIGEUR, and further certify that the prepared Bell-like states with $\theta {\text{ = }}{10^{\text{o}}}$ and $\theta {\text{ = }}{80^{\text{o}}}$ are steerable states. It means that two setting coherence steering criterion based on skew information of coherence can indeed detect more steerable states than the ones based on ${l_1}$ norm of coherence and relative entropy of coherence.

	\section{CONCLUSIONS}	
		In this work, we experimentally demonstrate the complementarity relations between quantum steering criteria by employing prepared Bell-like states with high fidelity and three Pauli operators. The experimental results are in accordance with the theoretical curves very well, and one can reveal the steerability of system by detecting the average coherence of subsystem. Whatever coherence measure is used, three measurement settings inequality is always obeyed by all prepared Bell-like states in experiment. Meanwhile, the experimental $S_0^B({\rho _{AB}})$ are anticorrelated with the $S_{12}^B({\rho _{AB}})/2$. If the prepared Bell-like states violate two setting coherence steering criteria, then the states cannot violate one setting coherence steering criteria. Furthermore, The strengths of coherence steering criteria rely on the choice of coherence measure. In comparison with two setting coherence steering criteria based on  ${l_1}$ norm of coherence and relative entropy of coherence, two setting coherence steering criterion based on skew information of coherence is more effective in witnessing steerable states. Our experimental results may deepen the understanding of the connection between the quantum steering and quantum coherence. 
		
	\section*{ACKNOWLEDGMENTS}	
		This work was supported by the National Science Foundation of China (Grants No. 11575001, No. 11405171, No. 61601002, and No. 11605028), Anhui Provincial Natural Science Foundation (Grant No. 2008085QA43), the Program of Excellent Youth Talent Project of the Education Department of Anhui Province of China (Grants No. gxyq2018059, and No. gxyqZD2019042, and No. gxyqZD2018065), the Natural Science Research Project of Education Department of Anhui Province of China (Grant No. KJ2018A0343), the Open Foundation for CAS Key Laboratory of Quantum Information (Grants No. KQI201801 and No. KQI201804). 
					
		Huan Yang and Zhi-Yong Ding contributed equally to this work.

	\bibliographystyle{plain}

\begin{thebibliography}{99}
		\bibitem{w01} A. Einstein, B. Podolsky, and N. Rosen, Can quantum-mechanical description of physical reality be considered complete? Phys. Rev. \textbf{47}, 777 (1935).
		\bibitem{w02} E. Schrödinger, Discussion of probability relations between separated systems, Math. Proc. Cambridge Philos. Soc. \textbf{31}, 555 (1935).
		\bibitem{w03} R. Horodecki, P. Horodecki, M. Horodecki, and K. Horodecki, Quantum entanglement, Rev. Mod. Phys \textbf{81}, 865 (2009).
		\bibitem{w04} J. S. Bell, On the Einstein Podolsky Rosen paradox, Physics \textbf{1}, 195 (1965).
		\bibitem{w05} N. Brunner, D. Cavalcanti, S. Pironio, V. Scarani, and S. Wehner, Bell nonlocality, Rev. Mod. Phys. \textbf{86}, 419 (2014).
		\bibitem{w06} H. M. Wiseman, S. J. Jones, and A. C. Doherty, Steering, Entanglement, Nonlocality, and the Einstein-Podolsky-Rosen Paradox, Phys. Rev. Lett. \textbf{98}, 140402 (2007).
		\bibitem{w07} R. Uola, A. C. S. Costa, H. C. Nguyen, and O. Gühne, Quantum steering, Rev. Mod. Phys. \textbf{92}, 015001 (2020).
		\bibitem{w08} E. G. Cavalcanti, S. J. Jones, H. M. Wiseman, and M. D. Reid, Experimental criteria for steering and the Einstein-Podolsky-Rosen paradox, Phys. Rev. A \textbf{80}, 032112 (2009).
		\bibitem{w09} A. C. S. Costa and R. M. Angelo, Quantification of Einstein-Podolsky-Rosen steering for two-qubit states, Phys. Rev. A \textbf{93}, 020103(R) (2016).
		\bibitem{w10} S. P. Walborn, A. Salles, R. M. Gomes, F. Toscano, and P. H. Souto Ribeiro, Revealing Hidden Einstein-Podolsky-Rosen Nonlocality, Phys. Rev. Lett. \textbf{106}, 130402 (2011).
		\bibitem{w11} J. Schneeloch, C. J. Broadbent, S. P. Walborn, E. G. Cavalcanti, and J. C. Howell, Einstein-Podolsky-Rosen steering inequalities from entropic uncertainty relations, Phys. Rev. A \textbf{87}, 062103 (2013).
		\bibitem{w12} A. C. S. Costa, R. Uola, and O. G\"uhne, Steering criteria from general entropic uncertainty relations, Phys. Rev. A \textbf{98}, 050104(R) (2018).
		\bibitem{w13} T Kriv\'achy, F Fr\"owis, and N Brunner, Tight steering inequalities from generalized entropic uncertainty relations, Phys. Rev. A \textbf{98}, 062111 (2018).
		\bibitem{w14} M. \.Zukowski, A. Dutta, and Z. Yin, Geometric Bell-like inequalities for steering, Phys. Rev. A \textbf{91}, 032107 (2015).
		\bibitem{w15} M. F. Pusey, Negativity and steering: A stronger Peres conjecture, Phys. Rev. A \textbf{88}, 032313 (2013).
		\bibitem{w16} Y. Z. Zhen, Y. L. Zheng, W. F. Cao, L. Li, Z. B. Chen, N. L. Liu, and K. Chen, Certifying Einstein-Podolsky-Rosen steering via the local uncertainty principle, Phys. Rev. A \textbf{93}, 012108 (2016).
		\bibitem{w17} D. J. Saunders, S. J. Jones, H. M. Wiseman, and G. J. Pryde, Experimental EPR-steering using Bell-local states, Nat. Phys. \textbf{6}, 845 (2010).
		\bibitem{w18} D. H. Smith, G. Gillett, M. P. de Almeida, C. Branciard, A. Fedrizzi, T. J. Weinhold, A. Lita, B. Calkins, T. Gerrits, H. M. Wiseman, S. W. Nam, and A. G. White, Conclusive quantum steering with superconducting transition-edge sensors, Nat. Commun. \textbf{3}, 625 (2012).
		\bibitem{w19} V. H\"andchen, T. Eberle, S. Steinlechner, A. Samblowski, T. Franz, R. F. Werner, and R. Schnabel, Observation of oneway Einstein-Podolsky-Rosen steering, Observation of one-way Einstein-Podolsky-Rosen steering, Nat. Photonics \textbf{6}, 596 (2012).
		\bibitem{w20} A. J. Bennet, D. A. Evans, D. J. Saunders, C. Branciard, E. G. Cavalcanti, H. M. Wiseman, and G. J. Pryde, Arbitrarily Loss-Tolerant Einstein-Podolsky-Rosen Steering Allowing a Demonstration over 1 km of Optical Fiber with No Detection Loophole, Phys. Rev. X \textbf{2}, 031003 (2012).
		\bibitem{w21} S. Armstrong, M.Wang, R. Y. Teh, Q. Gong, Q. He, J. Janousek, H. A. Bachor,M. D. Reid, and P. K. Lam, Multipartite Einstein-Podolsky-Rosen steering and genuine tripartite entanglement with optical networks, Nat. Phys. \textbf{11}, 167 (2015).
		\bibitem{w22} S. Wollmann, N. Walk, A. J. Bennet, H. M. Wiseman, and G. J. Pryde, Observation of Genuine One-Way Einstein-Podolsky-Rosen Steering, Phys. Rev. Lett. \textbf{116}, 160403 (2016).
		\bibitem{w23} K. Sun, X. J. Ye, J. S. Xu, X. Y. Xu, J. S. Tang, Y. C. Wu, J. L. Chen, C. F. Li, and G. C. Guo, Experimental Quantification of Asymmetric Einstein-Podolsky-Rosen Steering, Phys. Rev. Lett. \textbf{116}, 160404 (2016).
		\bibitem{w24} N. Tischler, F. Ghafari, T. J. Baker, S. Slussarenko, R. B. Patel, M. M. Weston, S. Wollmann, L. K. Shalm, V. B. Verma, S. W. Nam, H. C. Nguyen, H. M. Wiseman, and G. J. Pryde, Conclusive Experimental Demonstration of One-Way Einstein-Podolsky-Rosen Steering, Phys. Rev. Lett. \textbf{121}, 100401 (2018).
		\bibitem{w25} S. Wollmann, R. Uola, and A. C. S. Costa, Experimental demonstration of robust quantum steering, Phys. Rev. Lett. \textbf{125}, 020404 (2020).
		\bibitem{w26} H. Yang, Z-Y. Ding, D. Wang, H. Yuan, X-K. Song, J. Yang, C-J Zhang, and L. Ye, Experimental certification of the steering criterion based on a general entropic uncertainty relation, Phys. Rev. A \textbf{101}, 022324 (2020).
		\bibitem{w27} A. Streltsov, G. Adesso, and M. Plenio, Quantum coherence as a resource, Rev. Mod. Phys. \textbf{89}, 041003 (2017).
		\bibitem{w28} M. L. Hu, X. Y. Hu, J. C. Wang, Y. Peng, Y. R. Zhang, and H. Fan, Quantum coherence and geometric quantum discord, Phys. Rep. \textbf{762-764} 1 (2018).
		\bibitem{w29} T. Baumgratz, M. Cramer, and M. B. Plenio, Quantifying Coherence, Phys. Rev. Lett. \textbf{113} 140401 (2014).
		\bibitem{w30} C. Napoli, T. R. Bromley, M. Cianciaruso, M. Piani, N. Johnston, and G. Adesso, Robustness of Coherence: An Operational and Observable Measure of Quantum Coherence, Phys. Rev. Lett. \textbf{116}, 150502 (2016).
		\bibitem{w31} M. Piani, M. Cianciaruso, T. R. Bromley, C. Napoli, N. Johnston, and G. Adesso, Robustness of asymmetry and coherence of quantum states, Phys. Rev. A \textbf{93}, 042107 (2016).
		\bibitem{w32} S. Luo, Wigner-Yanase Skew Information and Uncertainty Relations, Phys. Rev. Lett. \textbf{91}, 180403 (2003).
		\bibitem{w33} D. Girolami, Observable Measure of Quantum Coherence in Finite Dimensional Systems, Phys. Rev. Lett. \textbf{113}, 170401 (2014).
		\bibitem{w34} C. S. Yu, Quantum coherence via skew information and its polygamy, Phys. Rev. A \textbf{95}, 042337 (2017).
		\bibitem{w35} A. Streltsov, U. Singh, H. S. Dhar, M. N. Bera, and G. Adesso, Measuring Quantum Coherence with Entanglement, Phys. Rev. Lett. \textbf{115}, 020403 (2015).
		\bibitem{w36} A. Winter, and D. Yang, Operational Resource Theory of Coherence, Phys. Rev. Lett. \textbf{116}, 120404 (2016).
		\bibitem{w37} T. R. Bromley, M. Cianciaruso, and G. Adesso, Frozen Quantum Coherence, Phys. Rev. Lett. \textbf{114}, 210401 (2015).
		\bibitem{w38} T. Theurer, D. Egloff, L. Zhang, and M. B. Plenio, Quantifying Operations with an Application to Coherence, Phys. Rev. Lett. \textbf{122}, 190405 (2019).
		\bibitem{w39} M-L. Hu, Y-Y. Gao, and H. Fan, Steered quantum coherence as a signature of quantum phase transitions in spin chains, Phys. Rev. A \textbf{101}, 032305 (2020).
		\bibitem{w40} Y. Yao, X. Xiao, L. Ge, and C. P. Sun, Quantum coherence in multipartite systems, Phys. Rev. A \textbf{92}, 022112 (2015).
		\bibitem{w41} K-D. Wu, Z-B. Hou, Y-Y. Zhao, G-Y. Xiang, C-F. Li, G-C. Guo, J-J. Ma, Q-Y. He, J. Thompson, and M. L. Gu, Experimental Cyclic Interconversion between Coherence and Quantum Correlations, Phys. Rev. Lett. \textbf{121}, 050401 (2018).
		\bibitem{w42} W. Zheng, Z. Ma, H. Wang, S-M. Fei, and X Peng, Experimental Demonstration of Observability and Operability of Robustness of Coherence, Phys. Rev. Lett. \textbf{120}, 230504 (2018).
		\bibitem{w43} A. \v Cernoch, K. Bartkiewicz, K. Lemr, and J. Soubusta, Experimental tests of coherence and entanglement conservation under unitary evolutions, Phys. Rev. A \textbf{97}, 042305 (2018).
		\bibitem{w44} Z-Y. Ding, H. Yang, D. Wang, H. Yuan, J. Yang, and L. Ye, Experimental investigation of entropic uncertainty relations and coherence uncertainty relations, Phys. Rev. A \textbf{101}, 032101 (2020).
		\bibitem{w45} U. Singh, M. N. Bera, H. S. Dhar, and A. K. Pati, Maximally coherent mixed states: Complementarity between maximal coherence and mixedness, Phys. Rev. A \textbf{91}, 052115 (2015).
		\bibitem{w46} S. Cheng and M. J. W. Hall1, Complementarity relations for quantum coherence, Phys. Rev. A \textbf{92}, 042101 (2015).
		\bibitem{w47} H-H. Qin, S-M. Fei, and X. Li-Jost, Trade-off relations of Bell violations among pairwise qubit systems, Phys. Rev. A \textbf{92}, 062339 (2015).
		\bibitem{w48} G. Sharma and A. K. Pati, Trade-off relation for coherence and disturbance, Phys. Rev. A \textbf{97}, 062308 (2018).
		\bibitem{w49} L-J. Zhao, L. Chen, Y-M. Guo, K. Wang, Y. Shen, and S-M. Fei, Trade-off relation among genuine three-qubit nonlocalities in four-qubit systems, Phys. Rev. A \textbf{100}, 052107 (2019).
		\bibitem{w50} D. Mondal, T. Pramanik, and A. K. Pati, Nonlocal advantage of quantum coherence, Phys. Rev. A \textbf{95}, 010301(R) (2017).
		\bibitem{w51} F. Pan, L. Qiu, and Z. Liu, The complementarity relations of quantum coherence in quantum information processing, Sci. Rep. \textbf{7}, 43919 (2017).
		\bibitem{w52} H. Zhu, Information complementarity: A new paradigm for decoding quantum incompatibility, Sci. Rep. \textbf{5}, 14317 (2015).
		\bibitem{w53} X. Zhan, X. Zhang, J. Li, Y. Zhang, B. C. Sanders, and P. Xue, Realization of the Contextuality-Nonlocality Tradeoff with a Qubit-Qutrit Photon Pair. Phys. Rev. Lett., \textbf{116} 090401 (2016).
		\bibitem{w54} M. M. Weston, M. J. W. Hall, M. S. Palsson, H. M. Wiseman, and G. J. Pryde, Experimental Test of Universal Complementarity Relations. Phys. Rev. Lett., \textbf{110} 220402 (2013).
		\bibitem{w55} W-M Lv, C Zhang, X-M Hu, H. Cao, J. Wang, and Y-F. Huang, Experimental test of the trade-off relation for quantum coherence, Phys. Rev. A \textbf{98}, 062337 (2018).
		\bibitem{w56} D. Mondal and D. Kaszlikowski, Complementarity relations between quantum steering criteria, Phys. Rev. A \textbf{98}, 052330 (2018).
		\bibitem{w57} P. G. Kwiat, E. Waks, A. G. White, I. Appelbaum, and P. H. Eberhard, Ultrabright source of polarization-entangled photons, Phys. Rev. A \textbf{60}, R773 (1999).
		\bibitem{w58} J. B. Altepeter, E. R. Jeffrey, and P. G. Kwiat, Adv. At. Mol. Opt. Phys. \textbf{52}, 105 (2005).
		\bibitem{w59} D. F. V. James, P. G. Kwiat, W. J. Munro, and A. G. White, Measurement of qubits, Phys. Rev. A \textbf{64}, 052312 (2001).
		\bibitem{w60} Z-B. Hou, J-F. Tang, C. Ferrie , G-Y Xiang, C-F. Li, and G-C Guo, Experimental realization of self-guided quantum process tomography, Phys. Rev. A \textbf{101}, 022317 (2020).
		\bibitem{w61} C. F. Li, J. S. Xu, X. Y. Xu, K. Li, and G.C. Guo, Experimental investigation of the entanglement-assisted entropic uncertainty principle, Nat. Phys. \textbf{7}, 752 (2011).	
	\end{thebibliography}
	
\end{document}